\documentclass[aip,reprint]{revtex4-1}
\usepackage{graphicx}
\usepackage{amsmath}
\usepackage{verbatim}

\begin{document}

\title{Electrostatic Charge Model for Dual-Layer Oxide Thin-Film Transistors
} 



\author{Måns J. Mattsson}
\email[Corresponding author: ]{mons.mattsson@gmail.com}
\affiliation{Department of Physics, Oregon State University, Corvallis, OR, 97331-6507, USA}

\author{John F. Wager}
\affiliation{School of Electrical Engineering and Computer Science, Oregon State University, Corvallis, OR, 97331-5501, USA}

\author{Matt W. Graham}
\affiliation{Department of Physics, Oregon State University, Corvallis, OR, 97331-6507, USA}


\date{\today}

\begin{abstract}
A simple electrostatic two-equation model for dual-layer thin-film transistor (TFT) operation is developed.  The model distributes electrostatic charge between the top and bottom semiconductor layers, and the resulting transfer and mobility curves accurately simulate experimental dual-layer a-IGZO/a-IZO TFT operation. The model further provides an analytic expression that maps charge confinement in the high-mobility a-IZO bottom semiconductor layer with the a-IGZO top-layer thickness and the conduction-band offset. By considering both a-IGZO/a-IZO layer charge partition and competing thickness-dependent oxygen vacancy trap density effects, the model suggests an optimal a-IGZO layer thickness of 9–12 nm. Importantly, this general electrostatic model extends to most dual-layer TFT systems and calculates how the top semiconductor layer TFT turn-on voltage changes sharply with the conduction band offset and layer thickness.



\end{abstract}

\maketitle 

Thin-film transistors (TFTs) using amorphous oxide semiconductors (AOS) as an active channel material have achieved widespread commercial success in the active-matrix flat-panel display industry. \cite{nomura2004room, kamiya2010present, geng2023thin,kim2023progress} Promising research efforts are further underway to redevelop AOS TFTs for next-generation memory devices, CMOS, and neuromorphic computing applications. \cite{ha2024exploring, belmonte2025disrupting, jang2022amorphous, yun2026neuromorphic,mattsson2025defect} In these applications, amorphous indium gallium zinc oxide (a-IGZO) is the dominant channel material due to its excellent device stability, low off current, and high electron mobility ($\sim$10--20~cm$^{2}$\,V$^{-1}$\,s$^{-1}$) compared to amorphous 
hydrogenated silicon, a-Si:H ($\sim$1~cm$^{2}$\,V$^{-1}$\,s$^{-1}$). \cite{stewart2016amorphous, zan2012achieving} Nonetheless, In-rich AOS materials, such as a-IZO and a-IGO, offer a much higher mobility range ($\sim$40--70~cm$^{2}$\,V$^{-1}$\,s$^{-1}$), but are seldom used as active channels because of significant device instability and bias-stress effects. \cite{zhang2025indium, hu2025tri, tang2025long, yuan2022stable} Dual-layer channel TFTs are therefore a promising compromise, in which a stabilizing layer, such as a-IGZO, is combined with a high-mobility layer, such as a-IZO, thereby offering a pathway toward stable, stress-resistant, high-mobility AOS TFTs that expand their potential use for high-mobility applications. \cite{wen2024high, kim2024improved, nahar2024study, nam2026effects, park2023high}
\par
In a dual-layer TFT, electron transport can occur in the top, bottom, or both semiconductor channel layers. Achieving high-mobility performance requires that electrons primarily accumulate in the high-mobility layer, while the low-mobility layer serves as a stabilizer. Still lacking is an intuitive physical model that analytically determines the dual-layer TFT charge distribution from basic material properties such as band offset and channel thicknesses. Numerical TCAD simulations can provide insight for a specific dual-layer TFT structure \cite{billah2021high,kim2022remarkable, huang2024enhancing}, but do not generalize to analytical expressions that provide an intuitive understanding of the dual-layer TFT system. The primary goal of this work is to develop and deploy a device physics-based electrostatic model for dual-layer TFT operation. The electrostatic model yields analytical expressions that define the conditions for when carrier accumulation is confined to the top or bottom semiconductor channel layer. Our model is applied to top-gated GI/a-IGZO/a-IZO TFTs and reveals that a maximum a-IGZO layer thickness of $\sim$12 nm is required to ensure carrier accumulation is confined to the higher-mobility bottom a-IZO layer.

\par
\begin{figure}[h]
    \centering
    \includegraphics[width=1 \linewidth]{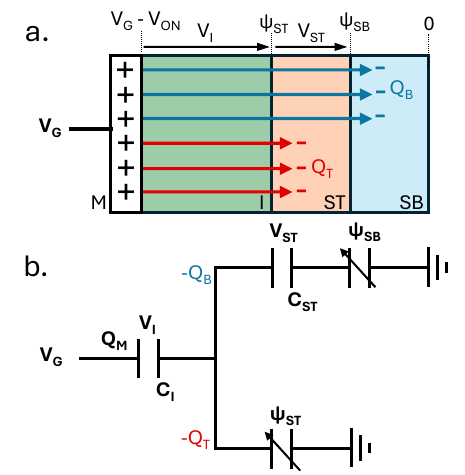}
 \caption{\textbf{(a)} Dual-layer TFT electrostatics. Under a forward (positive) gate overvoltage $(V_G - V_{\mathrm{ON}}$, where $V_{\mathrm{ON}}$ is the turn-on voltage), electric field lines originate on positive charge near the metal-insulator interface (Q$_M$) and terminate on negative charge (free or trapped) in the top (Q$_T$) or bottom (Q$_B$) semiconductor, giving rise to a voltage drop across the insulator (V$_I$) or semiconductors $(\psi_{ST}, V_{ST}, \psi_{SB})$. \textbf{(b)} Equivalent circuit of a dual-layer TFT, consisting of linear capacitors with voltage drops V$_I$ and V$_{ST}$, and non-linear voltage-controlled capacitors with voltage drops $\psi_{ST}$ and $\psi_{SB}$.}
    \label{fig:circuit}
\end{figure}

Figure \ref{fig:circuit}a illustrates charge balance in a dual-layer TFT in which positive charge at the metal-insulator interface (Q$_M$) is balanced by negative charge (free or trapped) in the top (Q$_T$) or bottom (Q$_B$) semiconductor. This charge distribution results in a voltage drop across the top ($\psi_{ST}$ and V$_{ST}$) and bottom ($\psi_{SB}$) of the semiconductor. Note that the voltage drop $\psi_{ST}$ arises from negative charge present in the top semiconductor, whereas V$_{ST}$ originates from negative charge present in the bottom semiconductor. Figure \ref{fig:circuit}b is the equivalent circuit of a dual-layer TFT. The voltage drops V$_I$ and V$_{ST}$ are modeled as linear capacitors, while the voltage drops $\psi_{ST}$ and $\psi_{SB}$ are represented as non-linear voltage-controlled capacitors.
\par
Dual-layer TFT electrostatic assessment follows from the Fundamental Equations included in Table I and involves the solution of seven equations [(i)-(vii)] with seven unknowns (V$_I$, V$_{ST}$, $\psi_{ST}$, $\psi_{SB}$, Q$_M$, Q$_T$, $Q_B$). Rearrangement of Eqs. (i)-(v) allows for simplification in terms of two equations,
\begin{align}
 V_G - V_{ON} &= -\frac{Q_{T}+Q_{B}}{C_I} + \psi_{ST},
    \label{Eq:model1} \\
     \psi_{ST} &= -\frac{Q_{B}}{C_{ST}} + \psi_{SB},
    \label{Eq:model2}
\end{align}
in two unknowns, $\psi_{ST}$ and $\psi_{SB}$, where Q$_T$ and Q$_B$ are evaluated using the Charge Density Equations included in Table I.
The two-equations-in-two-unknowns formulation of the dual-layer TFT electrostatics problem described differs significantly from an alternative assessment procedure recently proposed. \cite{wager2024dual} The key difference in this approach is that it is based on the formulation of an equivalent circuit, as mapped directly from charge balance electrostatics. \cite{wager2017device} The corresponding simple circuit analysis yields the electrostatic Eqs. (i)-(v) in Table I.
\begin{table}[h]
\centering
\footnotesize
\renewcommand{\arraystretch}{1.4}
\begin{tabular}{ll}
\multicolumn{2}{c}{Fundamental Equations:} \\

(i)   & $V_G - V_{\mathrm{ON}} = V_I + \psi_{ST}$ \\

(ii)  & $Q_M = -(Q_T + Q_B)$ \\

(iii) & $Q_M = C_I V_I$ \\

(iv)  & $Q_B = -C_{ST}V_{ST}$ \\

(v)   & $V_{ST} = \psi_{ST} - \psi_{SB}$ \\

(vi)  & $Q_T \equiv Q_T(\psi_{ST})$ \\

(vii) & $Q_B \equiv Q_B(\psi_{SB})$ \\

\multicolumn{2}{c}{Charge Density Equations:} \\

(I)   &
$Q_B(\psi_{SB})
= Q_{SB}(\psi_{SB}) + Q_{TSB}(\psi_{SB})$ \\

(Ia)  &
$Q_{SB}(\psi_{SB}) =
-\dfrac{\varepsilon_{SB}\sqrt{2k_B T}}
{qL_{D,SB}}
\sqrt{
\exp\!\left(
\dfrac{q\psi_{SB}}{k_B T}
\right)-1
}$ \\

(Ib)  &
$Q_{TSB}(\psi_{SB}) =
a_{SB}Q_{SB}
+
q\dfrac{b_{SB}}{2}
\left[
1+
\operatorname{erf}
\left(
\dfrac{\psi_{SB}}
{\sqrt{2}\psi_w}
\right)
\right]$ \\

(II)   &
$Q_T(\psi_{ST})
= Q_{ST}(\psi_{ST}) + Q_{TST}(\psi_{ST})$ \\

(IIa)  &
$Q_{ST}(\psi_{ST}) =
-\dfrac{\varepsilon_{ST}\sqrt{2k_B T}}
{qL_{D,ST}}
\sqrt{
\exp\!\left(
\dfrac{q(\psi_{ST}\!+\!\psi_{ST}(V_{ON})\!+\!\Delta E_C)}
{k_B T}
\right)-1
}$ \\

(IIb)  &
$Q_{TST}(\psi_{ST}) =
a_{ST}Q_{ST}
+
q\dfrac{b_{ST}}{2}
\left[
1+
\operatorname{erf}
\left(
\dfrac{\psi_{ST}\!+\!\psi_{ST}(V_{ON})\!+\!\Delta E_C}
{\sqrt{2}\psi_w}
\right)
\right]$ \\

\end{tabular}
\label{tab:fundamental_equations}
\caption{Dual-layer TFT electrostatic assessment is accomplished via solution of seven fundamental equations in seven unknowns (V$_I$, V$_{ST}$, $\psi_{ST}$, $\psi_{SB}$, Q$_M$, Q$_T$, $Q_B$: see Figure 1). Charge density equations account for both free (accumulation) charge (Q$_{ST}$ and Q$_{SB}$) and trapped charge (Q$_{TST}$ and Q$_{TSB}$). $\epsilon_{ST}$ and $\epsilon_{SB}$ is the permittivity of the top and bottom semiconductor layers, respectively. L$_{D,ST}$ and L$_{D,SB}$ is the effective Debye length of each semiconductor layer, determined by the short- or long-base formulation (see Refs.  \cite{wager2022thin, wager2023single}).}
\end{table}
\par
\begin{figure}[!htb]
    \centering
    \includegraphics[width=1 \linewidth]{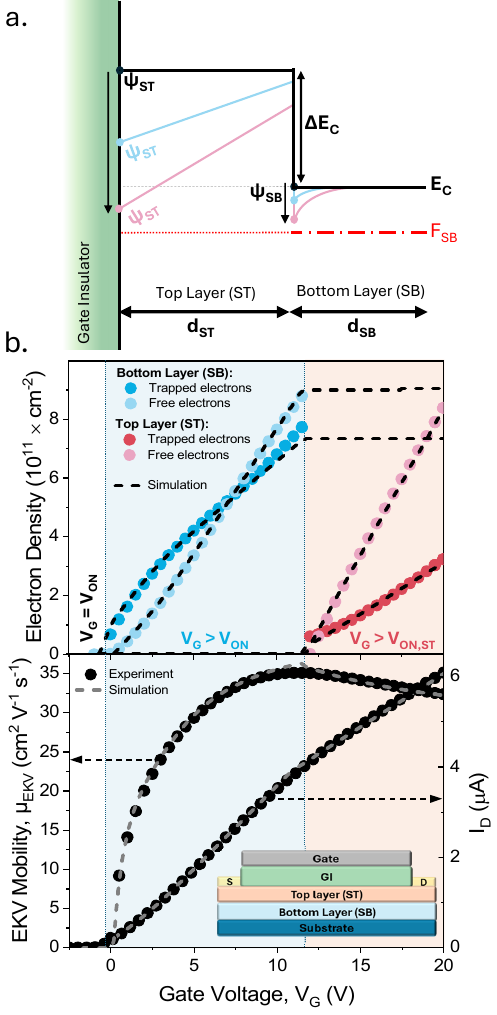}
 \caption{ \textbf{(a)} Energy band diagram for a dual-layer TFT for 3 different operating voltages: V$_G$ = 0 (black), V$_G > V_{ON}$ (blue), V$_G > V_{ON,ST}$ (red). \textbf{(b)} \textit{(top panel)} Simulated free and trapped electron densities in the top a-IGZO (red) and bottom a-IZO (blue) semiconductor layers for a GI/10 nm a-IGZO/7 nm a-IZO dual-layer TFT. \textit{(bottom panel)} The corresponding experimental (circles) and simulated (dashed lines) transfer and mobility curves. \textit{(inset)} Top-gated dual-layer TFT device schematic.}
    \label{fig:model}
\end{figure}

Before employing the developed dual-layer TFT Eqs. \ref{Eq:model1} and \ref{Eq:model2}, a few comments regarding the Charge Density Equations of Table I are pertinent. Equations (Ia) and (IIa) relate the free accumulation charge density to the representative surface potential for each semiconductor. The form of Eq. (Ia) is quite standard and is obtained by solving Poisson's equation for a single layer TFT,\cite{wager2024dual, wager2022thin, colinge2002physics} but the exponential term $\psi_{ST}(V_{ON})$+$\Delta E_C$ in Eq. (IIa) merits comment. As illustrated in Fig. \ref{fig:model}a, the close proximity of the bottom semiconductor conduction band minimum (E$_C$) and the quasi-Fermi level (F$_{SB}$) ensures that electron accumulation occurs abruptly as soon as $\psi_{SB}$ is greater than zero. By contrast, the onset of electron accumulation in the top semiconductor cannot occur until $\psi_{ST}$ is greater than $\psi_{ST}(V_{ON})$+$\Delta E_C$, the sum of the surface potential evaluated at V$_G$ = V$_{ON}$ of the top semiconductor and the conduction band discontinuity, $\Delta E_C$. This is because this onset of accumulation occurs when the conduction band minima of the top semiconductor reaches the same energetic position as the initial value (V$_G$ = V$_{ON}$) of the bottom semiconductor, and occurs when the total amount of surface band bending satisfies $\psi_{ST}$ = $\psi_{ST}(V_{ON})$+$\Delta E_C$.
\par
Note that the quasi-equilibrium approximation is employed in this analysis, such that F$_{SB}$ is flat, as it is extended across the top semiconductor, as indicated by the red dotted lines in Fig. \ref{fig:model}a. \cite{grove1967physics} Additionally, if the conduction bands are approximately flat at $V_G = V_{\mathrm{ON}}$, as is typically the case when turn-on is near zero ($V_{\mathrm{ON}} \approx 0$), the initial surface potential of the top semiconductor may be taken as negligible, i.e., $\psi_{\mathrm{ST}}(V_{ON}) = 0$. This assumption is adopted throughout the remainder of the paper. When this assumption of initially flat bands does not hold,  $\psi_{\mathrm{ST}}(V_{ON})$ instead behaves as a constant adjustment factor to $\Delta E_C$ (see Supplementary Material S6).
\par
Finally, Eqs. (Ib) and (IIb) in Table I account for electron trapping in the bottom and top semiconductor, respectively. Physically, the first term in each equation accounts for conduction band tail trapped charge, while the second term accounts for a Gaussian-shaped subgap trap of characteristic width $\psi_w$ (note that an error function is the integral of a Gaussian function).
\par
We are now ready to use the dual-layer TFT Eqs. \ref{Eq:model1} and \ref{Eq:model2} for device-physics assessment. The upper panel of Figure \ref{fig:model}b plots simulated free and trapped electron densities (black dashed) for both the top and bottom semiconductor layer of a GI/10 nm a-IGZO/7nm a-IZO TFT. Blue and red circles represent experimentally extracted electron densities for the bottom and top semiconductor channel layer, respectively. The simulation is in excellent agreement with the experimental results. As expected from the energy band diagram trend given in Fig. \ref{fig:model}a, the bottom a-IZO channel turns on abruptly near zero volts, while the top a-IGZO semiconductor channel onset occurs near 12 V. 
\par
The bottom panel of Figure \ref{fig:model}b plots the experimental and simulated transfer and EKV mobility \cite{wager2022amorphous} curves, showing excellent comparative agreement. Note how the transfer curve slope decreases appreciably in the red-shaded portion of the curve (where the lower-mobility top a-IGZO semiconductor turns on) compared to the blue-shaded portion of the curve (where carrier accumulation is confined to the high-mobility a-IZO semiconductor). More dramatically, the EKV mobility peaks at V$_G$ $\approx$ 12 V, corresponding to the top semiconductor turn-on voltage, and clearly distinguishes the gate voltage regimes in which electron accumulation occurs predominantly in the bottom or top semiconductor layer.
\par
The simulated drift mobility of a dual-layer TFT is calculated as,
\begin{equation}
\mu_{\mathrm{drift}}
=
\frac{
Q_{ST}\mu_{0ST}
+
Q_{SB}\mu_{0SB}
}
{
Q_{ST}+Q_{TST}+Q_{SB}+Q_{TSB}
}
\label{Eq:mobility2}
\end{equation}
where $\mu_{0ST}$ and $\mu_{0SB}$ are trap-free mobilities for top and bottom semiconductors, respectively. For the simulation included in Fig. \ref{fig:model}b bottom panel, $\mu_{0ST}$ = 36 cm$^{2}$V$^{-1}$s$^{-1}$ and $\mu_{0SB}$ = 66 cm$^{2}$V$^{-1}$s$^{-1}$ are needed to have the simulation match the experimental data. The $\mu_{0ST}$ value is unexpectedly large. Assuming thermally-limited diffusive transport,\cite{wager2022amorphous, wager2023single, wager2022thin}
\begin{equation}
    \mu_0=\frac{q\hbar}{6m^*k_BT}
\end{equation}
where q is the electronic charge, $\hbar$ is the reduced Planck constant, m$^*$ is electron effective mass, k$_B$ is Boltzmann's constant, and T is temperature. The effective mass for a-IGZO is commonly approximated to be 0.34 m$_0$, where m$_0$ is the electron rest mass, corresponding to $\mu_{0}$ = 22 cm$^{2}$V$^{-1}$s$^{-1}$. \cite{fung2009two, seino2025design, takagi2005carrier} Thus, the value $\mu_{0ST}$ = 36 cm$^{2}$V$^{-1}$s$^{-1}$ used for a-IGZO in our simulation is unexpectedly large. We observe a similar mobility enhancement in the measured field-effect mobility of the a-IGZO layer (as measured in the red shaded region in Fig. \ref{fig:model}b) in the dual-layer TFT ($\mu_{FE,ST}$ = 26 cm$^{2}$V$^{-1}$s$^{-1}$) compared to a single-layer TFT ($\mu_{FE,ST}$ = 17 cm$^{2}$V$^{-1}$s$^{-1}$). We ascribe this mobility enhancement to an increase in the average scattering time from 4.2 fs to about 6.7 fs. \cite{wager2022amorphous} Physically, this longer scattering time may be expected from a slight improvement in crystallinity when a-IGZO is deposited on a-IZO rather than SiO$_2$, thereby minimizing active channel disorder. Finally, assuming that thermally-limited diffusive transport is applicable, the value $\mu_{0SB}$ = 66 cm$^{2}$V$^{-1}$s$^{-1}$ suggests that m$^*$ $\approx$ 0.11 m$_0$ for electrons in a-IZO. Supplementary Material section S1 provides further details regarding the construction of Figure \ref{fig:model}b, including all parameters of the simulation and the experimental electron density extraction.

\begin{figure*}[!htb]
    \centering
    \includegraphics[width=1.0 \linewidth]{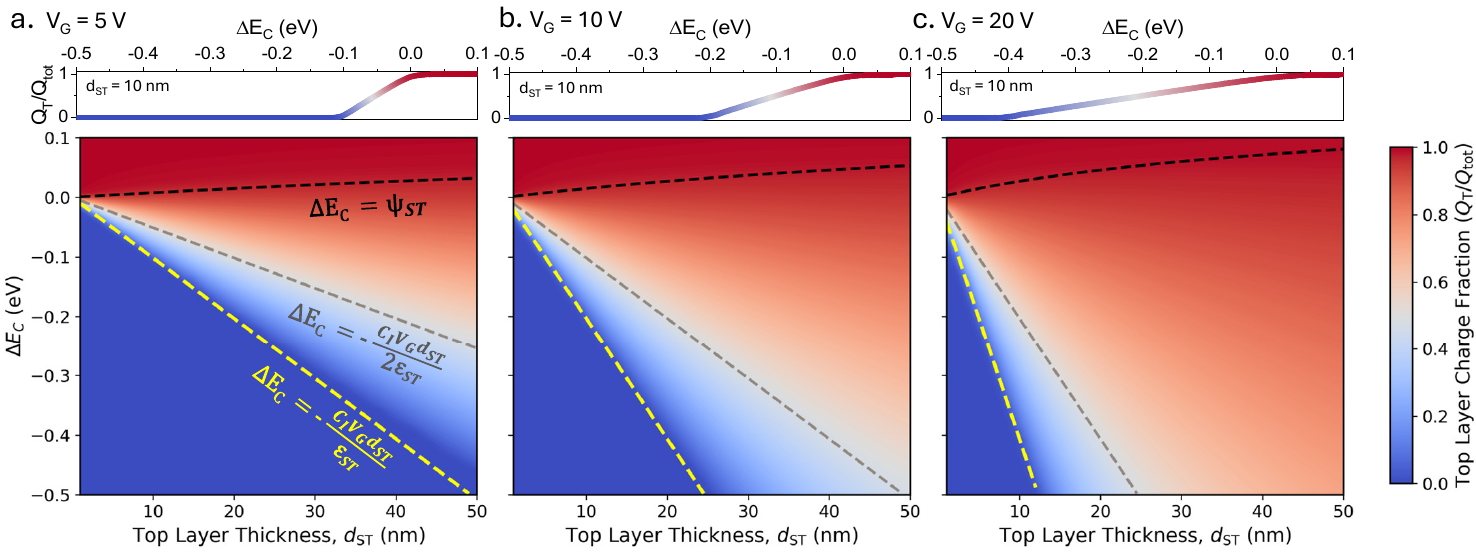}
 \caption{ Charge partition maps of the top-layer fractional charge (Q$_{T}$/Q$_{tot}$) as a function of conduction band offset, $\Delta E_C$, and top semiconductor layer channel thickness, d$_{ST}$, in a generalized top-gate dual-layer TFT under an applied gate overvoltage of \textbf{(a)} V$_G$ = 5 V, \textbf{(b)}  10 V and \textbf{(c)} 20 V. Upper panels show line cuts for d$_{ST}$ = 10 nm.}
    \label{fig:map}
\end{figure*}
\par
Figure \ref{fig:map} plots the simulated fraction of charge in the top semiconductor layer of a generalized dual-layer TFT. Blue or red regions of a charge partition map correspond to regimes of operation in which all of the accumulation charge resides in either the bottom or top semiconductor, respectively. White regions indicate equal charge partitioning, in which similar concentrations of accumulation charge are present in each semiconductor layer. Dashed lines denote analytical relations that identify charge-partition boundaries. These charge partition maps are insightful for optimizing the design of a top-gate dual-layer TFT, in terms of the conduction band discontinuity, $\Delta E_C$, and top semiconductor thickness, $d_{ST}$.
\par
Assuming that the bottom layer is a high-mobility semiconductor, an optimal dual-layer TFT is constrained to always operate in the dark blue region of the charge partition map, defined by $\Delta E_C \leq -C_I V_G d_{ST}/\epsilon_{ST}$. This ensures that electron conduction occurs exclusively in the high-mobility layer. In terms of conduction band discontinuity, a more negative value of $\Delta E_C$ is desirable to ensure carrier confinement in the high-mobility bottom semiconductor layer. Since conduction band discontinuity is defined as $\Delta E_C$ = E$_{C,SB}$ - E$_{C,ST}$, a negative value of $\Delta E_C$ means that the top semiconductor conduction band minimum is positioned above that of the bottom semiconductor. Ultimately, material constraints limit the practically achievable range of $\Delta E_C$. \cite{Liu2026UVOFGA, sanal2013growth} As discussed below, $\Delta E_C$ = -0.25 eV for the a-IGZO/a-IZO material system under consideration. Employing $\Delta E_C$ = -0.25 eV, it is clear from Figure \ref{fig:map} that the thickness of the top semiconductor layer establishes the maximum gate voltage, based on the constraint that TFT device operation be limited to dark blue regions of the charge partition map.  When operating a maximum gate voltage of V$_G$ = 10 V, Figure \ref{fig:map}b suggests that if $\Delta E_C$ = -0.25 eV, a maximum top layer thickness of d$_{ST}$ $\approx$ 12 nm is required to ensure that accumulation charge is confined in the high-mobility bottom-layer semiconductor.
\begin{figure}[!htb]
    \centering
    \includegraphics[width=1 \linewidth]{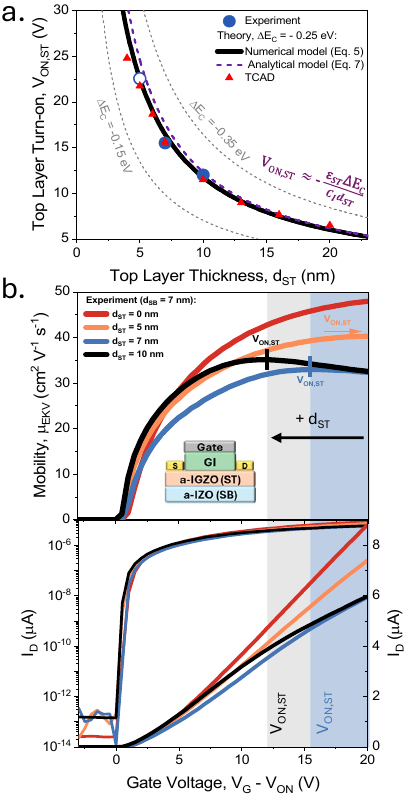}
 \caption{ \textbf{(a)} Top semiconductor layer turn-on voltage versus top semiconductor layer thickness for GI/a-IGZO/7nm a-IZO TFT. Experimental data (solid blue dot = measured; blue circle = extrapolated). Approximate analytical model (dashed). Numerical model (black solid) TCAD (red triangles). \textbf{(b)} Experimental EKV mobility (upper panel) and transfer curves (lower panel) for various a-IGZO top layer thicknesses. }
    \label{fig:comparison}
\end{figure}
\par
The analytical condition defining the boundary of the dark blue regions at which all accumulated charge resides in the bottom semiconductor layer (Q$_T$ = 0) and top semiconductor layer is at the onset of turn on ($\psi_{ST}$ = - $\Delta E_C$), corresponding to the yellow dashed line in Fig. \ref{fig:map}, can be derived by applying these conditions to Eqs. \ref{Eq:model1} and \ref{Eq:model2} which yields,
\begin{equation}
\Delta E_C =
-\frac{C_I}{C_I + C_{ST}} V_G
-
\frac{C_{ST}}{C_I + C_{ST}} \psi_{SB}
\label{blue}
\end{equation}
Equation \ref{blue} defines the boundary condition for charge confinement in the bottom semiconductor layer. However, the equation can only be solved by numerical iteration due to its explicit dependence upon $\psi_{SB}$. Recognizing that C$_{ST}$ $>>$ C$_I$ and that $\psi_{SB}$ is small compared to $|\Delta E_C|$, Eq. \ref{blue} can be approximated as,
\begin{equation}
    \Delta E_C \approx -\frac{C_I V_G d_{ST}}{\epsilon_{ST}},
    \quad \text{for } Q_T \rightarrow 0
    \label{approxblue}
\end{equation}
Equation \ref{approxblue} is used to draw the dashed yellow charge partition boundaries in Figure \ref{fig:map}. Finally, recognizing that Q$_T$ $\rightarrow$ 0 defines the onset of the top semiconductor layer turn-on voltage, V$_G$ = V$_{ON,ST}$, Eq. \ref{approxblue} can be rewritten as,
\begin{align}
   V_{ON,ST} &\approx -\frac{\epsilon_{ST} \Delta E_C}{C_I d_{ST}},
   \quad \text{for } Q_T \rightarrow 0
   \label{bluevg}
\end{align}
Equations \ref{blue}–\ref{bluevg} constitute the electrostatic theory results that describe the boundary condition where charge accumulation is confined to the bottom semiconductor layer, or equivalently, the gate voltage corresponding to the onset of top semiconductor layer accumulation. Derivations of the boundary relations based on electrostatic assessment for complete charge confinement in the top semiconductor and for the equal charge partition, corresponding to the dark red and white regimes in Fig. \ref{fig:map}, respectively, are provided in Supplementary Material Section S7.
\par
To test the validity of the developed electrostatic theory, Fig. \ref{fig:comparison}a compares against experimental a-IGZO/a-IZO dual-layer TFT data. Experimentally, V$_{ON,ST}$ is measured (closed blue circles) or estimated by extrapolation (open blue circle) when the turn on voltage is beyond the measurement range (see Supplementary Material S3). The analytical condition of Eq. \ref{bluevg} is plotted using $\Delta E_C$ = - 0.15, -0.25 and -0.35 eV, yielding the three dashed lines in Fig. \ref{fig:comparison}a. The purple dashed line, using $\Delta E_C$ = - 0.25 eV, shows good agreement with the three experimentally extracted values of V$_{ON,ST}$. For large V$_{ON,ST}$, a slightly better fit (black solid line, for $\Delta E_C$ = - 0.25 eV) to experimental data is obtained by numerical solution of Eq. \ref{blue}, as it represents the full electrostatic model without approximation. Finally, as a consistency check of the validity of the developed electrostatic theory, V$_{ON,ST}$ is extracted from TCAD simulated transfer curves (red triangles, for $\Delta E_C$ = -0.25 eV), matching both experiment and theory. 
\par
The conduction band discontinuity, $\Delta E_C$, is a critical device-physics parameter for modeling the electrical performance of a dual-layer TFT. Equation \ref{bluevg} enables a procedure for estimating $\Delta E_C$ directly from dual-layer TFT transfer curves by fitting to experimentally extracted V$_{ON,ST}$ values as a function of the top layer thickness, $d_{ST}$. Our estimate by this procedure of $\Delta E_C$ = -0.25 eV for a-IGZO/a-IZO is in excellent agreement with previously reported values of a-IGZO/a-IZO conduction band offset. \cite{billah2021high, kim2022remarkable, yang2025enhancing}
\par
\par
Figure \ref{fig:comparison}b presents experimental EKV mobility curves (upper panel) and corresponding transfer curves (lower panel) for various top semiconductor layer a-IGZO thicknesses, d$_{ST}$ = 10, 7, 5 and 0 nm. The mobility curves for the $d_{ST} = 10$ nm (black) and 7 nm (blue) devices peak due to the onset of electron accumulation in the a-IGZO layer at $V_G = V_{ON,ST}$. This transition corresponds to a change in the slope of the transfer curve and to a reduced current at large gate overvoltages compared to the single-layer (red) and $d_{ST} = 5$ nm devices, which exclusively conduct through the higher-mobility a-IZO layer. In contrast, varying the experimental thickness of the bottom a-IZO layer does not significantly affect $V_G = V_{ON,ST}$, consistent with the absence of $d_{SB}$ dependence of Eq. \ref{bluevg} (see Supplementary Material S4).
\begin{figure}[!htb]
    \centering
    \includegraphics[width=1 \linewidth]{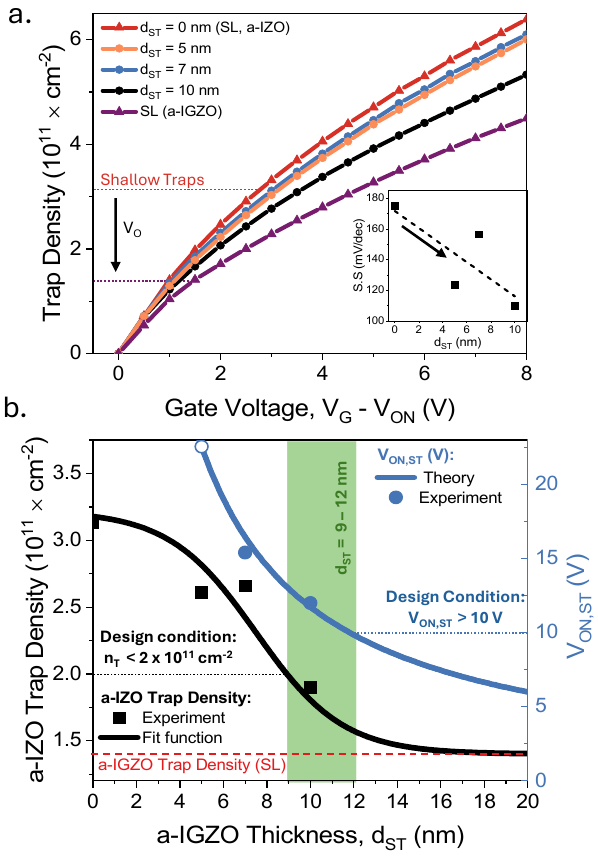}
 \caption{ \textbf{(a)} Extracted electron trap density as a function of gate overvoltage for single-layer a-IZO (red) and single-layer a-IGZO (purple) TFT, as well as for dual-layer a-IGZO/a-IZO TFTs with varying a-IGZO thickness, d$_{ST}$. \textbf{(b)} a-IZO layer trap density (black, left y-axis) and a-IGZO turn-on voltage (blue, right y-axis), V$_{ON,ST}$, as a function of a-IGZO top layer thickness. The green region highlights the optimal thickness range of d$_{ST}$ = 9-12 nm for minimizing trap density while confining electrons to the high-mobility a-IZO layer.}
    \label{fig:traps}
\end{figure}
\par
\par
Figure \ref{fig:traps}a presents accumulation trap density in the a-IZO bottom layer as extracted from experimental transfer curves for dual-layer TFTs with varying top layer thickness (Figure \ref{fig:comparison}b), with an addition for the trap density of a single-layer a-IGZO TFT. Since the trapped electron density at low gate overvoltage is dominated by shallow oxygen vacancy trapping, whereas at high overvoltage it is governed by conduction band tail trapping, the shallow oxygen vacancy concentration can be estimated from the total trap density at the point of slope change, which occurs approximately at the threshold voltage $V_G = V_{T}$. \cite{mattsson2026shallow} From Figure \ref{fig:traps}a, the estimated shallow oxygen vacancy concentration in the single layer a-IZO TFT (red) is $3.1\times 10^{11}$ cm$^{-2}$, whereas for the single layer a-IGZO TFT is significancy lower at $1.4\times 10^{11}$ cm$^{-2}$ (purple triangles). This lower shallow oxygen vacancy trap density in a-IGZO is consistent with the incorporation of Ga, which is known to increase the oxygen vacancy formation energy and thus suppress vacancy formation. \cite{kim2024improved,noh2011electronic,kim2012effects} Note from Figure \ref{fig:traps}a that the oxygen vacancy shallow trap concentration decreases with increasing a-IGZO thickness in a dual-layer TFT. These results suggest that the presence of a-IGZO protects the a-IZO surface from oxygen loss due to the harsh plasma environment associated with the deposition of the gate insulator.
\par
As shown in Figure \ref{fig:traps}a, compared to the single layer a-IZO TFT, the shallow oxygen vacancy trap density decreases with increasing a-IGZO thickness, reaching density of $1.9\times 10^{11}$ cm$^{-2}$ for d$_{ST}$ = 10 nm. This reduction in shallow trap density is reflected in the improved subthreshold swing, which decreases from 175 meV/dec in the SL a-IZO TFT to 110 meV/dec in the dual-layer a-IGZO/a-IZO TFT.  Since oxygen vacancies are believed to be responsible for numerous device instabilities and bias stress effects,\cite{mativenga2021origin,shin2025suppression,lee2025high} reducing their density while simultaneously confining electron accumulation to the high-mobility layer (in this case, a-IZO) may result in improved TFT performance. 
\par
Figure \ref{fig:traps}b illustrates the design trade-off between electron accumulation in the high-mobility layer and shallow trap density in GI/a-IGZO/a-IZO dual-layer TFTs. The blue line (right y-axis) plots V$_{ON,ST}$ for the a-IGZO top layer, while the black line (left y-axis) represents a best fit function for the estimated shallow trap densities (black squares) with varying a-IGZO thickness, d$_{ST}$. In order to confine electron accumulation to the high-mobility layer (in this case a-IZO) V$_{ON,ST}$ needs to be larger than the intended operating gate voltage. This is achieved by simply decreasing d$_{ST}$. However, decreasing $d_{ST}$ increases the a-IZO oxygen vacancy shallow trap density, potentially leading to increased instability and bias stress effects, as well as a degraded subthreshold swing. As a result, an optimal a-IGZO top layer thickness exists such that the shallow oxygen vacancy trap density is minimized, while electron accumulation is confined to the high-mobility a-IZO layer. For example, for an intended operating gate voltage range of 10 V (V$_{ON,ST}$ $>$ 10 V), and an upper acceptable limit of shallow oxygen vacancy trap density of $2\times 10^{11}$ cm$^{-2}$, the optimal a-IGZO top layer thickness range satisfying both criteria is $d_{ST} \approx 9$-$12$ nm.
\par
In summary, a device physics-based electrostatic model is developed for dual-layer TFT operation that reduces to a two-equation, two-unknown model described by Eqs. \ref{Eq:model1} and \ref{Eq:model2}. This simple model provides an analytic expression (Eq. \ref{bluevg}) that defines the boundary between the accumulation charge confinement to the bottom semiconductor layer and the onset of accumulation in the top semiconductor layer. The a-IGZO/a-IZO dual-layer TFT demonstrates the utility of this expression, showing that charge can be confined to the high-mobility a-IZO layer by tuning the a-IGZO thickness to maximize current drive over a given gate voltage range, while maintaining a lower shallow oxygen vacancy trap density than for a single-layer a-IZO TFT.  This device physics-based electrostatic modeling is a useful approach for designing, analyzing, and optimizing dual-layer TFTs. Moving forward, this electrostatic modeling approach can be extended to other dual-layer TFT systems, enabling selective charge confinement exclusively within the high-mobility semiconductor layer.
\par
\par
\section*{Supplementary Material}
Supplementary material, including simulation parameters, turn-on voltage extraction, surface potential correction, and numerical solutions, is provided as ancillary files with this arXiv submission.

\begin{acknowledgments}
This work is supported by a Samsung Global Research Outreach (GRO) Award and by a Samsung Electronics Co., Ltd. (IO250414-12603-01) grant.  Equipment is partly supported by a NSF Grant (DMR-1920368).  The authors acknowledge Applied Materials, Inc. for providing the a-IGZO/a-IZO dual-layer TFTs devices used in this study. 
\end{acknowledgments}

\section*{Conflict of Interest}
The authors have no conflicts to disclose.

\section*{Data Availability Statement}
The data that supports the findings of this study are available from the corresponding author upon request. 

%
%

%



\bibliography{mybib}

\end{document}